# Ultrafast ignition with relativistic shock waves induced by high power lasers

Shalom Eliezer [1], Noaz Nissim [2], Shirly Vinikman Pinhasi [2], Erez Raicher [2,3] and José Maria Martinez Val [1].

(1) Nuclear Fusion Institute, Polytechnic University of Madrid, Madrid, Spain.
(2) Applied Physics Division, Soreq NRC, Yavne, Israel.
(3) Hebrew University of Jerusalem, Jerusalem, Israel.

*6[th] April, 2014*

Abstract

In this paper we consider laser intensities larger than $10^{16}$ W/cm$^2$ where the ablation pressure is negligible in comparison with the radiation pressure. The radiation pressure is caused by the ponderomotive force acting mainly on the electrons that are separated from the ions to create a double layer (DL). This DL is accelerated into the target, like a piston that pushes the matter in such a way that a shock wave is created.

Here we discuss two novel ideas. First is the transition domain between the relativistic and non-relativistic laser induced shock waves. Our solution is based on relativistic hydrodynamics also for the above transition domain. The relativistic shock wave parameters, such as compression, pressure, shock wave and particle flow velocities, sound velocity and rarefaction wave velocity in the compressed target, and the temperature are calculated. Secondly, we would like to use this transition domain for shock wave induced ultrafast ignition of a pre-compressed target. The laser parameters for these purposes are calculated and the main advantages of this scheme are described. If this scheme is successful a new source of energy in large quantities may become feasible.

## 1. Introduction

The inertial fusion energy (IFE) is based on high compression [1-3]. The reasoning is that it is energetically cheaper to compress than to heat and the nuclear reactions are proportional to density square. IFE of deuterium-tritium (DT) requires high compression (> 1000) and in particular the aneutronic fusion [4-6] of the proton-boron11 needs extremely huge compression (>10,000). The high compression is achieved by shock waves and by the accumulation of matter during the stagnation of the implosion of the target shell.

The shock waves in laser plasma interaction [7] have played an important role in the study of inertial fusion energy. In 1974 the first direct observation of a laser-driven shock wave was reported [8]. A planar solid hydrogen target was irradiated with a 10 Joule, 5 ns, Nd laser and a pressure of about 2 Mbar was measured. Twenty years after this experiment, the Nova laser from Livermore created a pressure of $750\pm200$ Mbar [9]. This was achieved in a collision between two gold foils, where the flyer (Au foil) was accelerated by a high intensity x-ray flux created by the laser-plasma interaction.

In order to achieve nuclear fusion ignition, a mega-joule laser with few nanoseconds pulse duration has been constructed in USA [10]. The central spark ignition of DT is expected in the near future. In this scheme the target and the driver pulse shape are designed in such a way that only a spark at the center of the compressed fuel is heated and ignited [11,12]. The rest of the fuel is heated by α particles produced in the DT reactions.

In order to ignite a DT target with significantly less than few MJ of energy, it was suggested [13,14] to separate the drivers that compress and heat the target. This idea is called fast ignition (FI). FI triggers not in a central spark, but in a secondary interaction of an igniting driver of a very short duration, such as a multi PW laser beam. The PW laser is supposed to form a channel during a few picoseconds in the plasma atmosphere and to ignite a part of the fuel at the stagnation point of the implosion. For this purpose it was estimated that ignition requires about few tens of kJ of laser energy during duration of about 10 ps with irradiance of the order of $10^{20}$ W/cm$^2$. The FI problem is that the laser pulse does not penetrate directly into the compressed target with an electron density of the order of $10^{24}$ cm$^{-3}$. Therefore many schemes of FI were suggested: (1) the laser energy is converted into electrons that ignite the target [15], (2) the laser energy is converted into protons that ignite the target [16]. (3) Since the heating in the previous proposals is not confined and furthermore it is necessary to avoid preheating, a gold cone [17] (Au density/solid DT density ~ 100) was stuck in the spherical pellet. (4) FI is induced by plasma jets [18] that are induced by the same laser system that compresses the pellet. (5) The FI is done by plasma flow created from a thin exploding pusher foil [19, 20]. (6) Plasma blocks for FI were also suggested [21, 22]. (7) Murakami et al. [23] revived the old impact fusion with the help of the cone. (8) The use of clusters [24] was also suggested to ignite the compressed pellet. (9) Furthermore, it was suggested [25] to use an extra laser induced shock wave created by the same lasers that compressed the target in order to ignite the target. (10) Alternatively the fast ignition shock wave by a laser accelerated impact foil [26, 27] was proposed. The shock wave ignition schemes are actually based on heating by two shock wave collision using tailored laser pulses that were already suggested [28] even before the idea of fast ignition was explicitly published [13, 14].

It is well known that the interaction of a high power laser with a planar target creates a one dimensional (1D) shock wave [29, 30]. The theoretical basis for laser induced shock waves analyzed and measured experimentally so far is based on plasma ablation. For laser intensities $10^{12}$ W/cm$^2$ < $I_L$ < $10^{16}$ W/cm$^2$ and nanoseconds pulse duration hot plasma is created. This plasma exerts a high pressure on the surrounding material, leading to the formation of an intense shock wave moving into the interior of the target. The momentum of the out-flowing plasma balances the momentum imparted to the compressed medium behind the shock front similar to a rocket effect.

For $I_L$ < $10^{16}$ W/cm$^2$ the ablation pressure is dominant. For $I_L$ > $10^{16}$ W/cm$^2$ the radiation pressure is the dominant pressure at the solid-vacuum interface and the ablation pressure is negligible. In this last case the ponderomotive force drives the shock wave. For laser irradiances $I_L$ > $10^{21}$ W/cm$^2$ one gets a relativistic laser induced shock wave [31]. The theoretical foundation of relativistic shock waves is based on relativistic hydrodynamics [32] and was first analyzed by Taub [33]. Relativistic shock waves may be of importance in intense stellar explosions or in collisions of extremely high energy nuclear particles. Furthermore relativistic shock waves may be a new route for fast ignition nuclear fusion.

In section 2 the relativistic shock waves formalism is given for further consideration. In section 3 the laser induced shock wave equations are explicitly written and solved numerically without approximation for the first time. In the recent publication [31] the solution was given only for very strong relativistic shocks while in this paper the transition between the relativistic and nonrelativistic laser induced shock waves is obtained. It turns out that this transition domain is important and relevant for the fast ignition scheme as described in section 4. The paper is concluded with a short summary and discussion.

## 2. Relativistic shock waves.

The relativistic 1D (or non-relativistic [34]) shock wave is described by 5 variables: the particle density n (or the density $\rho = Mn$ where M is the particle mass), the pressure P, the energy density e, the shock wave velocity $u_s$ and the particle flow velocity $u_p$, assuming that we know the initial condition of the target: $n_0$ (or $\rho_0$), $P_0$, $e_0$ and, the particle flow velocity $u_0$, before the shock arrival. The 4 equations relating the shock wave variables are the 3 Hugoniot relations describing the conservation laws of energy, momentum and particle and the equation of state [35, 36] connecting the thermodynamic variables of the state under consideration. In order to solve the problem an extra equation is required and in our case we derive this equation from a laser-plasma interaction model.

The relativistic hydrodynamic starting point is the energy momentum 4-tensor $T_{\mu\nu}$ given by

$$T_{\mu\nu} = (e+P)U_\mu U_\nu + Pg_{\mu\nu} \qquad (1)$$

$U_\mu$ ($\mu = 0, 1, 2, 3$) is the dimensionless 4-velocity where the subscripts 0 is the time component and (1,2,3) are the space (x, y, z) components accordingly, and $g_{\mu\nu}$ is the metric tensor,

$$\mathbf{u}_\mu = cU_\mu = \left(\gamma c, \gamma \mathbf{v}_1, \gamma \mathbf{v}_2, \gamma \mathbf{v}_3\right),$$

$$g_{\mu\nu}: \; g_{00} = -1, g_{11} = g_{22} = g_{33} = 1, g_{\mu\nu} = 0 \text{ if } \mu \neq \nu, \qquad (2)$$

$$\gamma = \frac{1}{\sqrt{1-\beta^2}}; \;\; \beta = \frac{\mathbf{v}}{c}; \mathbf{v} = \sqrt{\mathbf{v}_1{}^2 + \mathbf{v}_2{}^2 + \mathbf{v}_3{}^2}.$$

where c is the speed of light in vacuum and v is the 3-dimension fluid particle velocity. Since our equation (1) is the starting point we write it more explicitly

$$T_{00} = \gamma^2\left(e+P\right) - P$$

$$T_{0i} = T_{i0} = -\gamma^2\left(e+P\right)\left(\frac{\mathbf{v}_i}{c}\right) \text{ for } i=1,2,3 \qquad (3)$$

$$T_{ij} = \gamma^2\left(e+P\right)\left(\frac{\mathbf{v}_i}{c}\right)\left(\frac{\mathbf{v}_j}{c}\right) + P\delta_{ij} \quad \text{for } i,j=1,2,3$$

In our 1D model one has for the velocity vector $\mathbf{v} = (v,0,0)$ and the Lorentz transformation is

$$\text{Lorentz transformation} = \begin{pmatrix} \gamma & -\gamma\beta & 0 & 0 \\ -\gamma\beta & \gamma & 0 & 0 \\ 0 & 0 & 1 & 0 \\ 0 & 0 & 0 & 1 \end{pmatrix} \qquad (4)$$

The energy-momentum conservation, the particle number conservation and the equation of state are given accordingly (Einstein summation is assumed from 0 to 3 for identical indexes)

$$\frac{\partial T_\mu{}^\nu}{\partial x^\nu} \equiv \partial_\nu T_\mu{}^\nu = 0 \text{ for } \mu=0,1,2,3.$$

$$\frac{\partial(nU^\mu)}{\partial x^\mu} \equiv \partial_\mu\left(nU^\mu\right) = 0 \qquad (5)$$

$$P = P\left(e,n\right)$$

We use equations (1), (3) and (5) for the conservation of energy density flux $c[T_{0x}]_0 = c[T_{0x}]_1$, the conservation of momentum density flux $[T_{xx}]_0 = [T_{xx}]_1$ and the conservation of particle number flux $[nU_x]_0 = [nU_x]_1$ along the shock wave singularity, with the subscripts 0 and 1 denoting accordingly the domains before and after shock arrival, to obtain the following equations

$$\gamma_0^2 \beta_0 \left(e_0 + P_0\right) = \gamma_1^2 \beta_1 \left(e_1 + P_1\right)$$
$$\gamma_0^2 \beta_0^2 \left(e_0 + P_0\right) + P_0 = \gamma_1^2 \beta_1^2 \left(e_1 + P_1\right) + P_1 \qquad (6)$$
$$\gamma_0 \beta_0 n_0 = \gamma_1 \beta_1 n_1$$

where $\gamma_i$ and $\beta_i = v_i/c$ for the domains 0 and 1 are defined in equation (2), where $v_0$ and $v_1$ are the inflow and outflow onto the shock wave singularity. Figure 1 describes the fluid flow velocities $v_0$ and $v_1$ as seen in the shock wave singularity frame of reference S and the shock wave velocity $u_{s1}$ and the particle flow velocities $u_{p1}$ and $u_{p0} = u_0$ as seen in the laboratory frame of reference.

From equations (6) the velocities $v_0$ and $v_1$ are obtained

$$\frac{v_0}{c} \equiv \beta_0 = \sqrt{\frac{\left(P_1 - P_0\right)\left(e_1 + P_0\right)}{\left(e_1 - e_0\right)\left(e_0 + P_1\right)}}$$
$$\frac{v_1}{c} \equiv \beta_1 = \sqrt{\frac{\left(P_1 - P_0\right)\left(e_0 + P_1\right)}{\left(e_1 - e_0\right)\left(e_1 + P_0\right)}} \qquad (7)$$

and the relativistic Hugoniot equation is derived [33],

$$\frac{\left(e_1 + P_1\right)^2}{n_1^2} - \frac{\left(e_0 + P_0\right)^2}{n_0^2} = \left(P_1 - P_0\right)\left[\frac{\left(e_0 + P_0\right)}{n_0^2} + \frac{\left(e_1 + P_1\right)}{n_1^2}\right] \qquad (8)$$

Assuming that in the laboratory the target is initially at rest, $u_0 = 0$, the shock wave velocity $u_s$ and the particle flow velocity $u_p$ in the laboratory frame of reference are related to the flow velocities $v_0$ and $v_1$ in the shock wave rest frame of reference by

$$u_s = -v_0$$
$$u_p = \frac{v_1 - v_0}{1 - \frac{v_0 v_1}{c^2}} \qquad (9)$$

The equation of state (EOS) taken here in order to calculate the shock wave parameters is the ideal gas EOS

$$e = \rho c^2 + \frac{P}{\Gamma - 1} \qquad (10)$$

where $\Gamma$ is the specific heat ratio and $v_0$ and $v_1$ are given in equations (7) .

### 3. Laser induced shock waves.

This paper analyzes the shock wave created in a planar target by the ponderomotive force induced by very high laser irradiance. In this domain of laser intensities the force acts on the electrons that are accelerated and the ions that follow accordingly. This model describes our piston model [37,38] as summarized schematically in figure 2: 2(a) the capacitor model for laser irradiances $I_L$ where the ponderomotive force dominates the interaction. 2(b) The system of the negative and positive layers is called a double layer (DL), $n_e$ and $n_i$ are the electron and ion densities accordingly, $E_x$ is the electric field, $\lambda_{DL}$ is the distance between the positive and negative DL charges, and $\delta$ is the solid density skin depth of the foil. The DL is geometrically followed by neutral plasma where the electric field decays within a skin depth and a shock wave is created. The shock wave description in the laboratory frame of reference is given in 2(c). This DL acts as a piston driving a shock wave [39, 40]. This model is supported in the literature by particle in cell (PIC) simulation [39, 41] and independently by hydrodynamic two fluid simulations [21, 22, 42]. The relativistic shock wave parameters, such as compression, pressure, shock wave and particle flow velocities, and temperature are calculated here for any compression $\kappa = \rho/\rho_0 > 1$ for the first time in the context of relativistic hydrodynamics. In a recent previous paper this was solved only for $\kappa = \rho/\rho_0 > 4$ with $\Gamma = 5/3$.

For $I_L < 10^{16}$ W/cm$^2$ the ablation pressure $P_a$ is dominant and it scales with the laser irradiance $I_L$ like $P_a \sim I_L^{\alpha}$ , where $\alpha$ is of the order of 2/3 in a 1D model [7]. For $I_L > 10^{16}$ W/cm$^2$ the radiation pressure is the dominant pressure at the solid-vacuum interface and the ablation pressure is negligible. In this last case the ponderomotive force drives the shock wave. The equations for relativistic hydrodynamics with the ideal gas equations of state (EOS) in the laboratory frame of reference are

$$(i) \quad \frac{u_{p1}}{c} = \sqrt{\frac{(P_1 - P_0)(e_1 - e_0)}{(e_0 + P_1)(e_1 + P_0)}}$$

$$(ii) \quad \frac{u_{s1}}{c} = \sqrt{\frac{(P_1 - P_0)(e_1 + P_0)}{(e_1 - e_0)(e_0 + P_1)}}$$

$$(iii) \quad \frac{(e_1 + P_1)^2}{\rho_1^2} - \frac{(e_0 + P_0)^2}{\rho_0^2} = (P_1 - P_0)\left[\frac{(e_0 + P_0)}{\rho_0^2} + \frac{(e_1 + P_1)}{\rho_1^2}\right]$$

$$\left.\begin{matrix}(iv)\\(v)\end{matrix}\right\} e_j = \rho_j c^2 + \frac{P_j}{\Gamma - 1}; \; j{=}0,1.$$

(11)

We have to solve these 5 equations together with our piston model equation [31, 38]

$$(vi) P_1 = \frac{2I_L}{c}\left(\frac{1-\beta}{1+\beta}\right); \beta \equiv \frac{u_{p1}}{c}$$

(12)

(11) and (12) describe 6 equations with 6 unknowns: $u_s$, $u_{p1}$, $P_1$, $\rho_1$, $e_1$ and $e_0$ assuming that we know $I_L$, $\rho_0$, $P_0$, $\Gamma$ and $u_o=0$. We take ideal gas EOS with $\Gamma = 5/3$. The calculations are conveniently done in the dimensionless units defined by

$$\Pi_L \equiv \frac{I_L}{\rho_0 c^3}; \kappa \equiv \frac{\rho_1}{\rho_0}; \kappa_0 \equiv \frac{\Gamma+1}{\Gamma-1}; \Pi \equiv \frac{P_1}{\rho_0 c^2}; \Pi_0 \equiv \frac{P_0}{\rho_0 c^2}; \qquad (13)$$

It is important to emphasize that if we take $P_0 = 0$ then we get only the $\kappa > \kappa_0$ solutions [31], therefore in order to see the behavior at the transition between relativistic and nonrelativistic domain one has to take $P_0 \neq 0$! In our numerical estimations we take $P_0 = 1$ bar $=10^6$ in cgs units. For example, the Hugoniot equation (11)$_{(iii)}$ together with the EOS equations (11)$_{(iv)+(v)}$ yield

$$\frac{P_0}{P_1} = \frac{\Pi_0}{\Pi} = \mathbf{0} \Rightarrow \begin{cases} \Pi = -B\left(\Pi_0 = 0\right) = \frac{\left(\Gamma-1\right)^2}{\Gamma}\kappa\left(\kappa-\kappa_0\right) \\ \kappa \equiv \frac{\rho_1}{\rho_0} \geq \kappa_0 \end{cases} \qquad (14)$$

$$\frac{P_0}{P_1} = \frac{\Pi_0}{\Pi} \neq \mathbf{0} \Rightarrow \begin{cases} \Pi^2 + B\Pi + C = 0 \\ \kappa \equiv \frac{\rho_1}{\rho_0} \geq 1 \end{cases}$$

$$\Pi = \left(\frac{1}{2}\right)\left(-B \pm \sqrt{B^2 - 4C}\right) \qquad (15)$$

$$B = \frac{\left(\Gamma-1\right)^2}{\Gamma}\left(\kappa_0\kappa - \kappa^2\right) + \Pi_0\left(\Gamma-1\right)\left(1-\kappa^2\right)$$

$$C = \frac{\left(\Gamma-1\right)^2}{\Gamma}\left(\kappa - \kappa_0\kappa^2\right)\Pi_0 - \kappa^2\Pi_0^{\;2}$$

The compression $\kappa$ as a function of the dimensionless pressure $\Pi = P/(\rho_0 c^2)$ is given in figure 3 for $\kappa_0 = 4$ ($\Gamma = 5/3$). Although $P_0/P_1$ is extremely small one cannot neglect it in the very near vicinity of $\kappa_0$ and in this domain one has to solve numerically equation (15). Furthermore, in order to see the transition between the relativistic and nonrelativistic approximation (see appendix A) one has to solve the relativistic equations with (15) in order to see the transition effects like the one shown in figure 3. However for $\kappa > \kappa_0$, for $(\kappa-\kappa_0)/\kappa_0 > 10^{-3}$, the approximation of equation (14) is very good in calculating the shock wave variables as a function of the dimensionless laser irradiance $\Pi_L$.

The numerical solutions of equations (11) and (12) are shown in figures 4 and 5. Figure 4 gives the dimensionless shock wave pressure $\Pi = P/(\rho_0 c^2)$ versus the dimensionless laser irradiance $\Pi_L = I_L/(\rho_0 c^3)$ in the domain $10^{-4} < \Pi_L < 1$. For a better understanding of this graph and for the practical proposal in the next section, the inserted table shows numerical values in the area $10^{-4} < \Pi_L < 10^{-2}$. Figure 5 describes the dimensionless shock wave velocity $u_s/c$ and the particle velocity $u_p/c$ in the laboratory frame of reference versus the dimensionless laser irradiance $\Pi_L = I_L/(\rho_0 c^3)$ in the domain $10^{-4} < \Pi_L < 1$, while the inserted table shows numerical values in the area $10^{-4} < \Pi_L < 10^{-2}$. As a numerical example we take a target (liquid deuterium-tritium (DT)) with initial density $\rho_0 = 0.2$ g/cm$^3$ irradiated by a laser with intensity $I_L = 5\times10^{22}$ W/cm$^2$, namely $\Pi_L = 9.26\times10^{-2}$. In this case our relativistic equations yield a compression $\kappa = \rho/\rho_0 = 4.09$, a pressure $P = 2\times10^{13}$ bars, a shock wave velocity $u_s = 0.35c$ and a particle velocity $u_p = 0.27c$ where c is the speed of light.

The relativistic speed of sound $c_S$ for an ideal gas EOS is

$$\frac{c_s}{c} = \sqrt{\left(\frac{\partial P}{\partial e}\right)_S} = \left(\frac{\Gamma P}{e+P}\right)^{1/2} = \left[\frac{\Gamma(\Gamma-1)\Pi}{\Gamma\Pi+(\Gamma-1)\kappa}\right]^{1/2} \tag{16}$$

In the shocked medium the characteristic velocity of a disturbance from the piston to the shock wave front, equal to the rarefaction wave in the shocked medium $c_{rw}$, is given by

$$c_{rw} = \frac{c_S + u_p}{1+\left(\dfrac{c_S u_p}{c^2}\right)} \tag{17}$$

Figure 6a and 6b describe accordingly the speed of sound in units of speed of light, $c_S/c$, and the ratio of shock velocity to the rarefaction velocity, $u_s/c_{rw}$, as a function of the dimensionless laser irradiance $\Pi_L = I_L/(\rho_0 c^3)$ in the domain $10^{-4} < \Pi_L < 1$. The inserted tables show numerical values in the area $10^{-4} < \Pi_L < 10^{-2}$.

We analyze now the temperature problem. The partial pressures of an ideal gas that contains electrons and ions with appropriate densities $n_e$, $n_i$ and temperature $T_e$,$T_i$ are $P_e$ and $P_i$ that can be described by

$$P_e = n_e k_B T_e ; P_i = n_i k_B T_i \tag{18}$$

If the associated photons in this system are in thermal equilibrium then a radiation temperature $T_r$ can be defined with a radiation pressure $P_r$ given by [35]

$$P_r = (1/3)aT_r^4 ; a = \left(\frac{1}{15}\right)\left(\frac{k_B^4}{h^3 c^3}\right) = 7.56\cdot10^{-15}[erg/(cm^3 K^4)]. \tag{19}$$

For a plasma in local thermal equilibrium satisfying $T_e = T_i = T_r = T$ where the ions have an ionization Z and an atomic number A, implying a ion mass of $Am_p$ where $m_p$ is the proton mass, the plasma pressure is given by

$$P = P_i + P_e + P_r = (Z+1)n_i k_B T + \left(\frac{1}{3}\right) a T^4 \qquad (20)$$

If the ion density satisfies

$$n_i [cm^{-3}] << 1.56 \times 10^{27} \left(\frac{k_B T}{m_e c^2}\right) \qquad (21)$$

then the radiation pressure is dominant and the temperature is given by

$$T \approx \left(\frac{3P}{a}\right)^{1/4} \qquad (22)$$

It is conceivable to assume that electrons and ions are in thermal equilibrium, i.e. $T_e = T_i$, however the shocked area is not optically thick for the energetic photons. In this case the energetic photons created by bremsstrahlung are leaving the system, implying $T_r << T_e$ or one can have a situation where radiation temperature is not defined at all. Therefore if the photon radiation in equation (19) is negligible one has

$$k_B T = m_p c^2 \left(\frac{A}{Z+1}\right)\left(\frac{\Pi}{\kappa}\right) \qquad (23)$$

Therefore in general we can write that the plasma temperature is constrained in the following domain

$$\left(\frac{m_p c^2}{k_B}\right)\left(\frac{A}{Z+1}\right)\left(\frac{\Pi}{\kappa}\right) > T > \left(\frac{3P}{a}\right)^{1/4} \qquad (24)$$

Taking the example given above for liquid DT with A = 2.5, Z = 1, $m_p$ = 938.3 MeV/c$^2$, initial density $\rho_0$ = 0.2 g/cm$^3$ irradiated by a laser with intensity $I_L$ = 5×10$^{22}$ W/cm$^2$, namely $\Pi_L$ = 9.26×10$^{-2}$ we get $\Pi$ = 0.11, $\kappa$ =4.09 and a temperature in the domain 26.2 keV < $k_B T$ < 31.6 MeV. However, for $k_B T$ >1 MeV we have electron-positron pair production [45, 46] and new physics is required here for the temperature calculations. It is out of the scope of this paper to analyze here this exotic case.

### 4. An ultrafast ignition solution to the energy problem.

In order to solve the energy problem of all generations the scientists have considered using controlled nuclear fusion energy. One of these approaches is the well-known inertial confinement fusion driven by high power lasers where the physics is based on compressing and igniting rather than confining the fuel [1, 2]. In order to ignite the fuel with less energy it was suggested to separate the drivers that compress and ignite the target [13, 14]. First the fuel is compressed then a second driver ignites a small part of the fuel while the created alpha particles in the deuterium-tritium (DT) interaction heat the rest of the target. This idea is called fast ignition. The fast ignition problem is that the laser pulse does not penetrate directly into the compressed target; therefore many schemes have been suggested [43].

The laser-solution of the energy problem requires very sophisticated High Power Lasers Science and Engineering (HPLSE). In a recent paper [44] the various HPLSE optimizations and design constrains for a laser fusion power plant are beautifully summarized and analyzed. From the many possible proposals to solve the energy problem with high power lasers (HPL) we consider 3 criteria for choosing the best candidate (present or future): (i) *Understanding the physics*. In HPL-target interaction there are many scientific problems not yet fully understood, such as laser-plasma instabilities, hydrodynamic instabilities, equations of state, non-linear transport issues, non-local thermodynamic equilibrium, etc., without neglecting the energy conservation! (ii) *Engineering simplicity*. The inertial fusion energy (IFE) project is extremely complicated technologically and therefore a major effort is required in choosing the laser system, the target design, etc., from all possible proposal by the physicists. The technological simplicity must be seriously taken into account. For example, IFE requires about $10^8$ or more laser shots per year; therefore complicated target designs (like inserting a golden cone inside a pellet) are not realistic. (iii) Last but not least IFE is supposed to be *economically practical.* This implies the required gain, defined as the nuclear energy output divided by the laser input per shot to be larger than 100 and furthermore the cost of a target should not be more expensive than about 0.1 US $.

Taking into account these 3 criteria it looks that (a) direct drive is simpler than indirect drive. (b) Fast ignition needs significantly less energy (about 0.3MJ instead of 3 MJ). Therefore the direct drive fast ignition has the potential to be the best route to achieve nuclear fusion as an energy source. (c) From all presently known fast ignition schemes the simplest fast ignition seems to be by an "extra shock" wave. We suggest a novel shock wave ignition scheme with less energy (in comparison with the present shock wave ignition scheme [25]) and without laser-plasma instabilities (no more than $I_L \lambda_L^2 = 10^{14} (W/cm^2) \mu m^2$ in the laser compression pulses). In this proposal the ignition shock wave is created by high irradiance laser and the shock wave is induced by ponderomotive force in the intermediate domain between the relativistic and non-relativistic hydrodynamics. For this case the relativistic shock wave formalism has to be considered as developed in our previous section. We call our scheme ultrafast since the laser pulse duration for the ignition process is significantly is smaller by one to two orders of magnitude.

The shock wave ignition criteria for DT nuclear fuel are

$$(i) \quad "\rho R" = \kappa \rho_0 \left( u_s - u_p \right) \tau_L \geq 0.3 [g/cm^2]$$

$$(ii) \quad T \geq 10 keV$$

(25)

For the DT fusion one has A = 2.5 and Z = 1, therefore equation (23) for 10 keV temperatures and a compression of $\kappa = 4$ implies a minimum dimensionless pressure $\Pi_{min} = 3.4 \times 10^{-5}$. According to our solution the dimensionless laser irradiance satisfies $\Pi_L > \Pi_{L,min} = 1.8 \times 10^{-5}$. $\Pi_{min}$ and $\kappa = 4$ infer a minimum shock velocity and particle velocity $u_s/c = 0.59 \times 10^{-2}$ and $u_p/c = 0.44 \times 10^{-2}$ accordingly. Using these values in equation (25)$_{(i)}$, one gets a laser pulse duration of $\tau_L = 1.6$ ps. Assuming a pre-compression of $\rho_0 = 10^3$ g/cm$^3$ the $\Pi_{L,min} = 1.8 \times 10^{-5}$ requires $I_L = 4.8 \times 10^{22}$ W/cm$^2$. The shock wave thickness turn out to be $l_s = (u_s - u_p)\tau_L = 0.72$ μm. In order to have a 1D shock wave to a reasonable approximation we require a laser focal spot radius $R_L = 5l_s$ implying a laser cross section of $S = \pi R_L^2 = 4.0 \times 10^{-7}$ cm$^2$. In this case the laser energy $W_L$ and power $P_L$ are 30kJ and 19 PW accordingly. This example was taken to describe our concept in figure 7. As a numerical example in this figure we take an initial pellet with radius $R_0 = 1$mm and DT fuel of density 0.2 g/cm$^3$ with thickness 0.1 mm (i.e. an aspect ratio of 10) that is compressed to a density of $\rho_0 = 10^3$ g/cm$^3$ (with a radius of 67 μm) by the nanosecond lasers. The picosecond fast igniter laser with a 7.2 μm in diameter creates a shock wave pulse with a thickness of 0.72 μm can be consider a 1D shock wave to a reasonable good approximation. The compressed pellet has a radius much larger than $\sqrt{S} \gg l_s$ in order to have a 1D shock wave. In table 1 we show how larger values of $\Pi_L$ change the laser and shock wave parameters.

The compression of a typical pellet as discussed in the literature [12, 47] requires between 100 to 300 kJ of energy depending on the equations of state, target design and the final required density. The fast ignition in our case needs about 30kJ of energy. Such a laser is under development and may be available in the near future.

## 5. Summary and discussion.

Recently [31] it was suggested that relativistic shock waves with shock wave velocity of 50% the speed of light and more can be created in the laboratory with high power lasers that are recently under development. In this paper we discuss two novel ideas. First is the transition domain between the relativistic and non-relativistic laser induced shock waves. Secondly, we would like to use this transition domain for shock wave induced ultrafast ignition of a pre-compressed target. The laser parameters for these purposes are calculated and the main advantages of this scheme are described. The many laser beams with the few nanoseconds pulse that compresses the target do not require $I_L \lambda_L^2 = 10^{15}$ W/cm$^2$µm$^2$ as in the previously proposed shock wave ignition scheme [25], thus the disturbing laser plasma instabilities do not occur. Furthermore in the present scheme less energy is required in the main laser pulses where a picosecond laser with very high power (~30PW) is required for the ultrafast ignition with the shock wave in the intermediate domain between the relativistic and non-relativistic hydrodynamics.

The present existing Petawatt lasers (see appendix B) might be used to start relativistic experimental research in the laboratory. The recent and future developments of high power lasers in the multi Petawatt domain could be important for relativistic shock waves in the laboratory with pressures of $10^{15}$ atmospheres or energy densities of the order of $10^{14}$ J/cm$^3$. Such pressures or energy densities have been suggested so far only in astrophysical objects.

The ultrafast ignition scheme suggested in this paper appears advantageous in comparison with the many fast ignition proposals, as given in our introduction section. It is based on the following merit of credit criteria: (i) *Understanding the physics*, (ii) *Engineering simplicity* and (iii) *are economically practical.* We think that the shock wave fast ignition is the best choice and the model suggested here between the relativistic and non-relativistic domain has significant advantages and should be taken seriously into account.

The solution suggested in this paper, like all other solutions to the energy problem is extremely scientifically difficult, lot of money and enormous optimism is required for a positive solution. The High Power Lasers Science and Engineering is complex, complicated but possible. If civilization is to survive we need new large quantities of energy. To quote Mark Twain (1835-1910): "And what is a man without energy? Nothing-nothing at all"

## Appendix A

For convenience we write the nonrelativistic Hugoniot equations and the ideal gas EOS:

$$
\begin{aligned}
&(i)\ u_{p1} = \left[P_1 - P_0\right]^{1/2} \left(\frac{1}{\rho_0} - \frac{1}{\rho_1}\right)^{1/2} \\[2em]
&(ii)\ u_s = \left(\frac{1}{\rho_0}\right) \frac{\left[P_1 - P_0\right]^{1/2}}{\left(\dfrac{1}{\rho_0} - \dfrac{1}{\rho_1}\right)^{1/2}} \\[2em]
&(iii)\ E_1 - E_0 = \left(\frac{1}{2}\right)\left[P_1 + P_0\right]\left(\frac{1}{\rho_0} - \frac{1}{\rho_1}\right) \\[2em]
&\left.\begin{array}{c}(iv)\\(v)\end{array}\right\}\ E_j = \left(\frac{1}{\Gamma - 1}\right)\left(\frac{P_j}{\rho_j}\right)\ \text{for j=0,1}
\end{aligned}
\tag{26}
$$

The equations are obtained from the relativistic equations (11) by using $e = \rho c^2 + \rho E$, $P$ and $\rho E$ are much smaller than $\rho c^2$ and $v/c \ll 1$.

## Appendix B

In this appendix we give a list of Petawatt lasers that are in use in different laboratories as for the end of the year 2013. The following data is not officially confirmed however this was used in literature and conferences according to our knowledge.

USA

Michigan University, Ann Arbor: 10J/30fs

Texas University, Austin:　　　 186J/167fs

Berkeley National Laboratory:　 40J/40fs

Rochester University, Rochester:　1kJ/1ps

LLNL, Livermore:　　　　　　 600J/500fs

CHINA

Beijing National Laboratory:　　　　　　 32J/28fs

Shanghai Institute of Optics and Fine Mechanics: 35J/27fs

EUROPE

Central Laser Facility, UK:  500J/500fs & 15J/30fs

Jena, Germany:　　　　　 120J/120fs

GSI Darmstadt, Germany:　 500J/500fs

<u>JAPAN</u>
Osaka University: 500J/500fs
<u>S. KOREA</u>
Gwangju University:  34J/30fs

**References**

1. J. H. Nuckolls, L. Wood, A. Thiessen, and G. B. Zimmermann. Nature **239**, 139 (1972).

2. G. Velarde and N. Carpintero-Santamaria eds. in *Inertial Confinement Nuclear Fusion: a Historical Approach by its Pioneers* (Foxwell and Davies Pub., UK, 2007).

3. K. Mima, M. Murakami, S. Nakai, and S. Eliezer, in *Applications of Laser-Plasma Interactions,* S. Eliezer, and K. Mima., eds. (CRC Press, Boca Raton, 2009).

4. J. M. Martinez Val, S. Eliezer, M. Piera, and G. Velarde. Phys. Lett. A**216**, 142 (1996).

5. S. Eliezer, and J. M. Martinez Val. Laser Particle Beams, **16**, 581 (1998)

6. S. Son, and N. J. Fish. Phys. Lett. A**329**, 76 (2004).

7. S. Eliezer, *The Interaction of High Power Lasers with Plasmas* (CRC Press, Boca Raton, 2002).

8. C. G. M. Van Kessel, and R. Sigel. *Phys. Rev. Lett.* **33,** 1020, (1974).

9. R. Cauble, D. W. Phillion, T. J. Hoover, N. C. Holmes, J. D. Kilkenny, and R. W. Lee. *Phys. Rev. Lett.* **70,** 2102 (1993).

10. E. I. Moses. *Nucl. Fusion* **49,** 104022 (2009).

11. J. D. Lindl. *Inertial Confinement Fusion: The Quest for Ignition and High Gain Using Indirect Drive.* (Springer, New York, 1997).

12. S. Atzeni and J. M. ter-Vehn. *The Physics of Inertial Fusion,* (Clarendon Press, Oxford, 2004).

13. N. G. Basov, S. Y. Guskov, and L. P. Feoktistov. J. Sov. Laser Res. **13**, 396 (1992).

14. M. Tabak, J. Hammer, M. E. Glinsky, W. L. Kruer, S. C. Wilks, J. Woodworth, E. M. Campbell, M. D. Perry, and R. J. Mason. Phys. Plasmas **1**, 1626 (1994).

15. P. A. Norreys, R. Allot, R. J. Clarke, J. Colliers, D. Neely, S. J. Rose, M. Zepf, M. Santala, A. R. Bell, K. Krushelnick, A. E. Dangor, N. C. Woolsey, R. G. Evans, H. Habara, T. Norimatsu, and R. Kodama. Phys. Plasmas **7,** 3721 (2000).

16. M. Roth, E. T. Cowan, M. H. Key, S. P. Hatchett, C. Brown, W. Fountain, J. Johnson, D. M. Pennington, R. A. Snavely, S. C. Wilks, K. Yasuike, H. Ruhl, F. Pegoraro, S. V. Bulanov, E. M. Campbell, M. D. Perry, and H. Powell. Phys. Rev. Lett. **86,** 436 (2001).

17. R. Kodama, P. A. Norreys, K. Mima, A. E. Dangor, R. G. Evans, H. Fujita, Y. Kitagawa, K. Krushelnick, T. Miyakoshi, N. Miyanaga, T. Norimatsu, S. J. Rose, T. Shozaki, K. Shigemori, A. Sunahara, M. Tampo, K. A. Tanaka, Y. Toyama, and M. Zepf. Nature **412**, 798 (2001).

18. J. M. Martinez Val, and M. Piera. Fusion Tech. **32,** 131 (1997).

19. A. Caruso, and C. Strangio. Laser and Particle Beams **19,** 295 (2001).

20. S. Y. Guskov. Quantum Electronics **31**, 885 (2001).

21. P. Lalousis, I. B. Foldes, and H. Hora. Laser and Particle Beams **30**, 233 (2012).

22. P. Lalousis, H. Hora, S. Eliezer, J. M. Martinez Val, S. Moustaizis, G. H. Miley, and G. Mourou. Phys. Lett. A **377**, 885 (2013).

23. M. Murakami, H. Nagatomo, H. Azechi, F. Ogando, M. Perlado, and S. Eliezer. Nucl. Fusion **46,** 9 (2006).

1717

24. S. Eliezer, J. M. Martinez Val, and C. Deutsch. Laser and Particle Beams **13**, 43 (1995).
25. R. Betti, C. D. Zhou, K. S. Anderson, L. J. Perkins, W. Theobald, and A. A. Sokolov. Phys. Rev. Lett. **98**, 155001 (2007).
26. S. Eliezer, and J. M. Martinez Val. Laser and Particle Beams **29**, 175 (2011).
27. S. Eliezer and S. V. Pinhasi. High Power Laser Science and Engineering **1,** 44 (2013).
28. S. Jackel, D. Saltzmann, A. Krumbein, and S. Eliezer. Phys. Plasma **26**, 3138 (1983).
29. V. E. Fortov, and I. V. Lomonosov. Shock Waves **20,** 53 (2010).
30. S. Eliezer, in *Laser-Plasma Interactions and Applications,* 68[th] Scottish Universities Summer School in Physics, eds. P. D. McKenna P, D. Neely, R. Bingham, and D. A. Jaroszynski, (Springer Publication, Heidelberg, pp 49-78 2013).
31. S. Eliezer, N. Nissim, E. Raicher, and J. M. Martinez Val. Laser and Particle Beams (2014), doi:10.1017/S0263034614000056.
32. L. D. Landau and E. M. Lifshitz. *Fluid Mechanics*, 2[nd] Edition. (Pergamon Press. Oxford, 1987).
33. A. H. Taub. Phys Rev, **74**, 3 (1948).
34. Y. B. Zeldovich and Y. P. Raizer. *Physics of Shock Waves and High Temperature Hydrodynamic Phenomena.*( Academic Press  Publications, New York, 1966).
35. S. Eliezer, A. Ghatak, and H. Hora. *Fundamental of Equation of State* (World Scientific, Singapore, 2002).
36. S. Eliezer and R. A. Ricci eds. *High Pressure Equation of State: Theory and Application,* Enrico Fermi School CXIII 1989 (North Holland Pub., Amsterdam, 1991).
37. S. Eliezer, and H. Hora. Double layers in laser produced plasmas, Physics Reports **172,** 339 (1989).
38. S. Eliezer, N. Nissim, J. M. Martinez Val, K. Mima, and H. Hora. Laser and Particle Beam (2014), doi:10.1017/S0263034613001018.
39. N. Naumova, T. Schlegel, V. T. Tikhonchuk., C. Labaune, I. V. Sokolov,  and G. Mourou. Phys. Rev. Le*tters* **102,** 025002 (2009).
40. S. Eliezer, J. M. Martinez Val, and S. V. Pinhasi. Laser and Particle Bea*ms* **31,** 113 (2013).
41. T. Esirkepov,  M. Borghesi, S. V. Bulanov., G. Mourou,  and T. Tajima. *Phys Rev Lett*., **92**, 175003 (2004).
42. H. Hora, P. Lalousis, and S. Eliezer.  Phys. Rev Letters **53**, 1650 (1984).
43. S. Y. Guskov. Plasma Physics Reports **39,** 1 (2013).
44. S. E. Bodner, A. J. Schmitt, and J. D. Sethian. High Power Laser Science and Engineering **1,** 2 (2013).
45. A. Di Piazza, C. Muller, K. Z. Hatsagortsyan, and C. H. Keitel. Reviews of Modern Physics **84**, 1177 (2012).
46. R. Ruffini, G. Vereshchagin, and S, Xue. Physics Reports **487**, 1 (2010).

47. S. Eliezer, M. Murakami, and J. M. Martinez Val. Laser and Particle Beams **25,** 585 (2007).

## TABLES

| $\Pi_L$ | $\rho_0[g/cm^3]$ | $I_L[W/cm^2]$ | $\kappa$ | $(u_s-u_p)/c$ | $\tau_L[ps]$ | $l_s[\mu m]$ | $S[cm^2]$ | $W_L[kJ]$ | $P_L[PW]$ |
|---------|------------------|---------------|----------|----------------|--------------|--------------|-----------|-----------|-----------|
| $1.8\times10^{-5}$ | $10^3$ | $4.8\times10^{22}$ | 4 | $0.15\times10^{-2}$ | 1.6 | 0.72 | $4.0\times10^{-7}$ | 30 | 19 |
| $1\times10^{-4}$ | $10^3$ | $2.7\times10^{23}$ | 4 | $0.5\times10^{-2}$ | 0.5 | 0.75 | $4.4\times10^{-7}$ | 60 | 120 |
| $1\times10^{-3}$ | $10^3$ | $2.7\times10^{24}$ | 4 | $1.3\times10^{-2}$ | 0.2 | 0.78 | $4.8\times10^{-7}$ | 260 | 1300 |

<u>Table 1</u>: The laser is defined by its irradiance $I_L$, pulse duration $\tau_L$, energy $W_L$ and power $P_L$. This laser creates a shock wave with a compression $\kappa$ in a pre-compressed target with initial density $\rho_0$. The shock wave thickness (= $(u_s-u_p)\tau_L$; $u_s$ and $u_p$ are the shock wave velocity and the particle velocity accordingly) and its cross section are $l_s$ and S appropriately satisfy $\sqrt{S} \gg l_s$ in order to have a 1D shock wave.

**Figure captions**

<u>Figure 1:</u> The fluid flow velocities $v_0$ and $v_1$ as seen in the shock wave singularity frame of reference S and the shock wave velocity $u_{s1}$ and the particle flow velocities $u_{p1}$ and $u_{p0} = u_0$ as seen in the laboratory frame of reference.

<u>Figure 2:</u> (a) The capacitor model for laser irradiances $I_L$ where the ponderomotive force dominates the interaction. (b) The parameters that define our capacitor model: $n_e$ and $n_i$ are the electron and ion densities accordingly, $E_x$ is the electric field, $\lambda_{DL}$ is the distance between the positive and negative DL charges. The DL is geometrically followed by neutral plasma where the electric field decays within a skin depth $\delta$ and a shock wave is created. (c) The shock wave description in the piston model.

<u>Figure 3:</u> The compression $\kappa = \rho/\rho_0$ as a function of the shock wave dimensionless pressure $\Pi = P/(\rho_0 c^2)$. The numerical values are obtained for $\Gamma = 5/3$.

<u>Figure 4:</u> The dimensionless shock wave pressure $\Pi = P/(\rho_0 c^2)$ versus the dimensionless laser irradiance $\Pi_L = I_L/(\rho_0 c^3)$ in the domain $10^{-4} < \Pi_L < 1$. For a better understanding of this graph the inserted table show numerical values in the area $10^{-4} < \Pi_L < 10^{-2}$.

<u>Figure 5:</u> The dimensionless shock wave velocity $u_s/c$ and the particle velocity $u_p/c$ in the laboratory frame of reference versus the dimensionless laser irradiance $\Pi_L = I_L/(\rho_0 c^3)$ in the domain $10^{-4} < \Pi_L < 1$. For a better understanding of this graph the inserted table show numerical values in the area $10^{-4} < \Pi_L < 10^{-2}$.

<u>Figure 6:</u> The speed of sound $c_S$ is given in units of the speed of light c in (a) and the ratio of the shock velocity to the rarefaction velocity, $u_s/c_{rw}$ is shown in (b) as function of the dimensionless laser irradiance $\Pi_L = I_L/(\rho_0 c^3)$ in the domain $10^{-4} < \Pi_L < 1$. The inserted tables show numerical values in the area $10^{-4} < \Pi_L < 10^{-2}$.

<u>Figure 7:</u> The fast ignition scheme suggested in this paper. As a numerical example an initial pellet with radius $R_0 = 1$mm and DT fuel of density 0.2 g/cm$^3$ with thickness 0.1 mm (i.e. an aspect ratio of 10) is compressed to a density of $\rho_0 = 10^3$ g/cm$^3$ by the nanosecond lasers with a radius of 67 μm. The picosecond fast igniter laser with a 7.2 μm in diameter creates a shock wave pulse with a thickness of 0.72 μm can be consider a 1D shock wave to a reasonable approximation.

Figure 1:

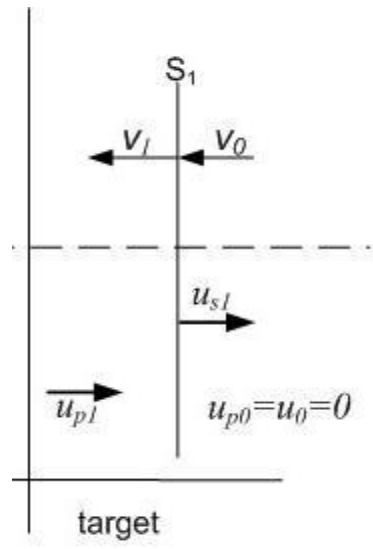

target

Figure 2:

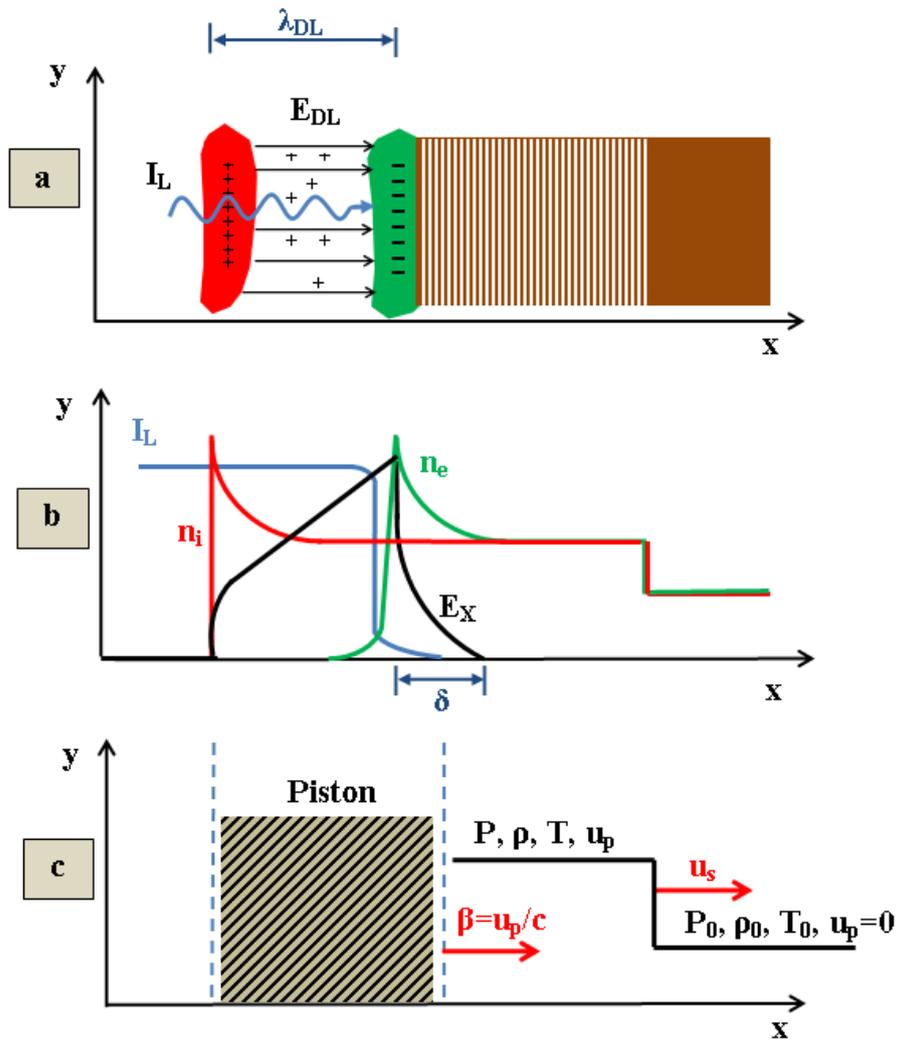

Figure 3:

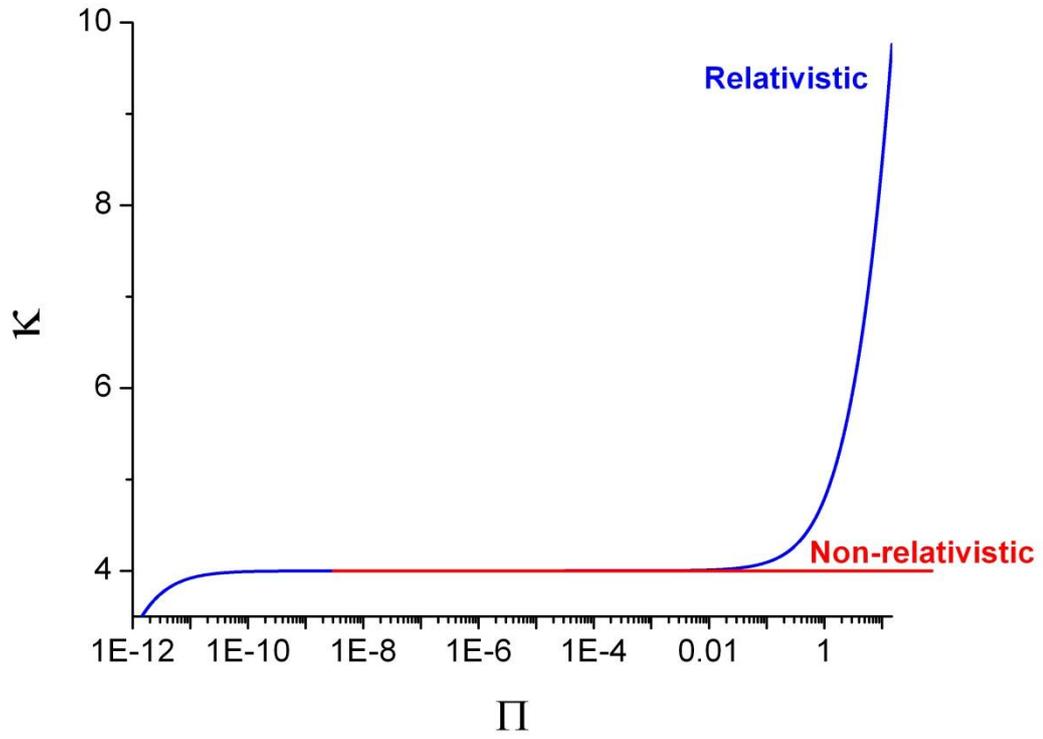

Figure 4:

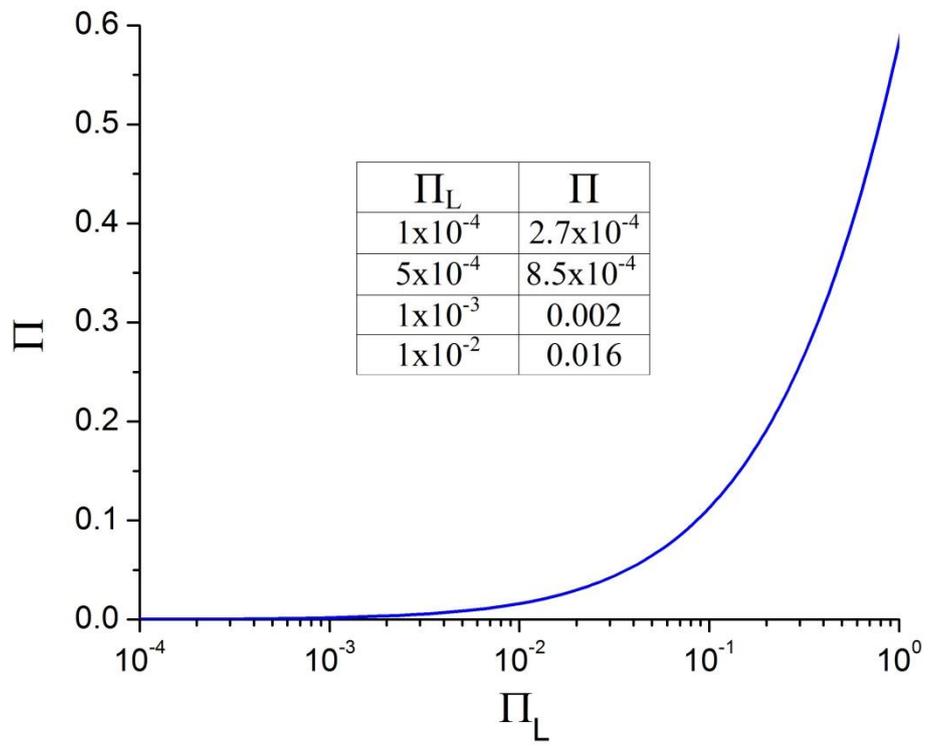

Figure 5:

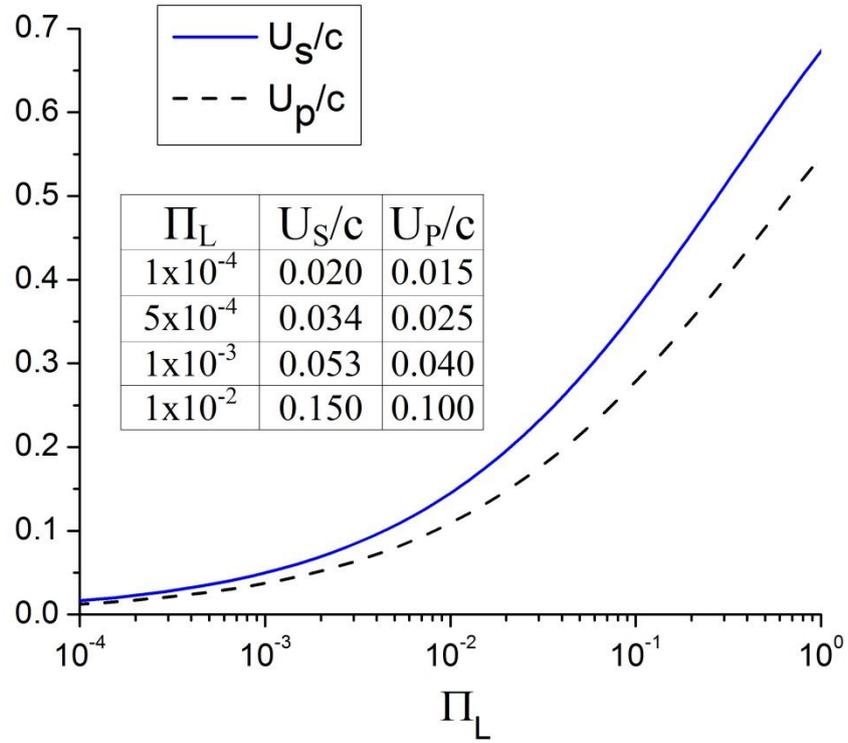

Figure 6:

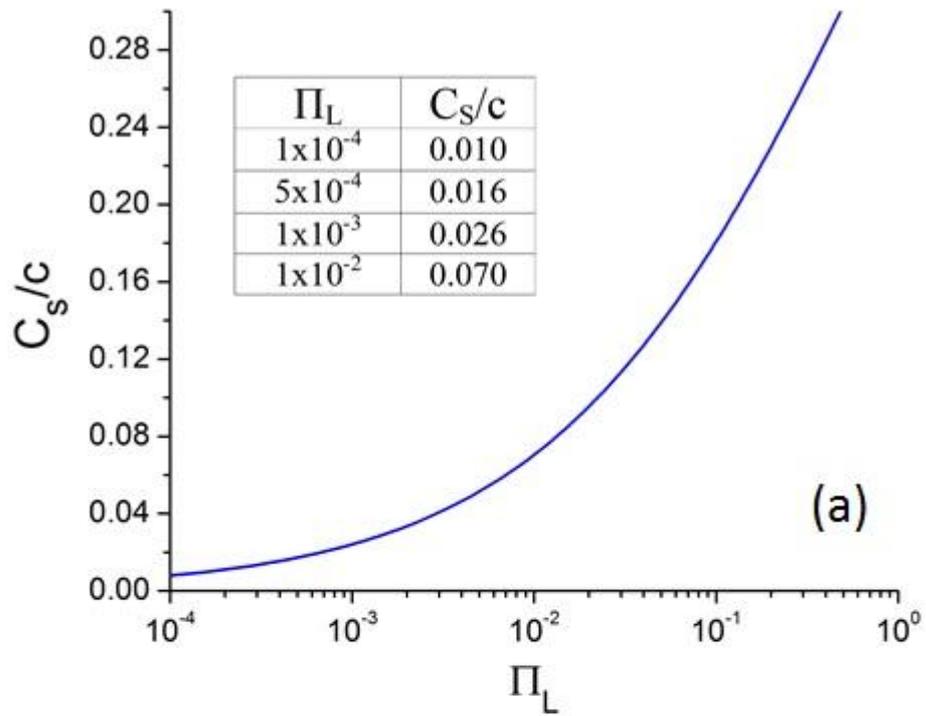

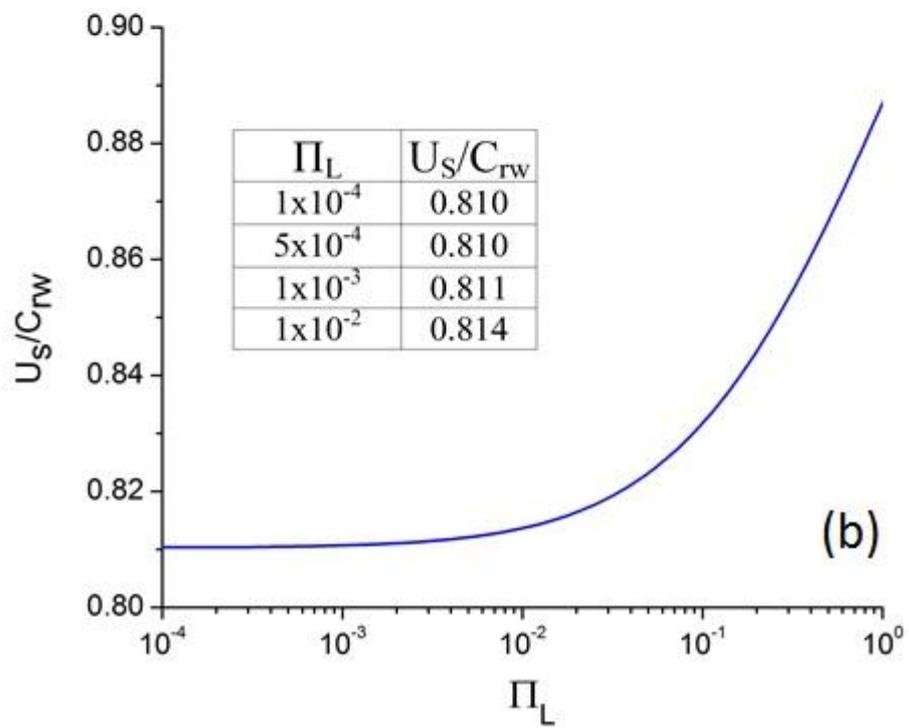

Figure 7:

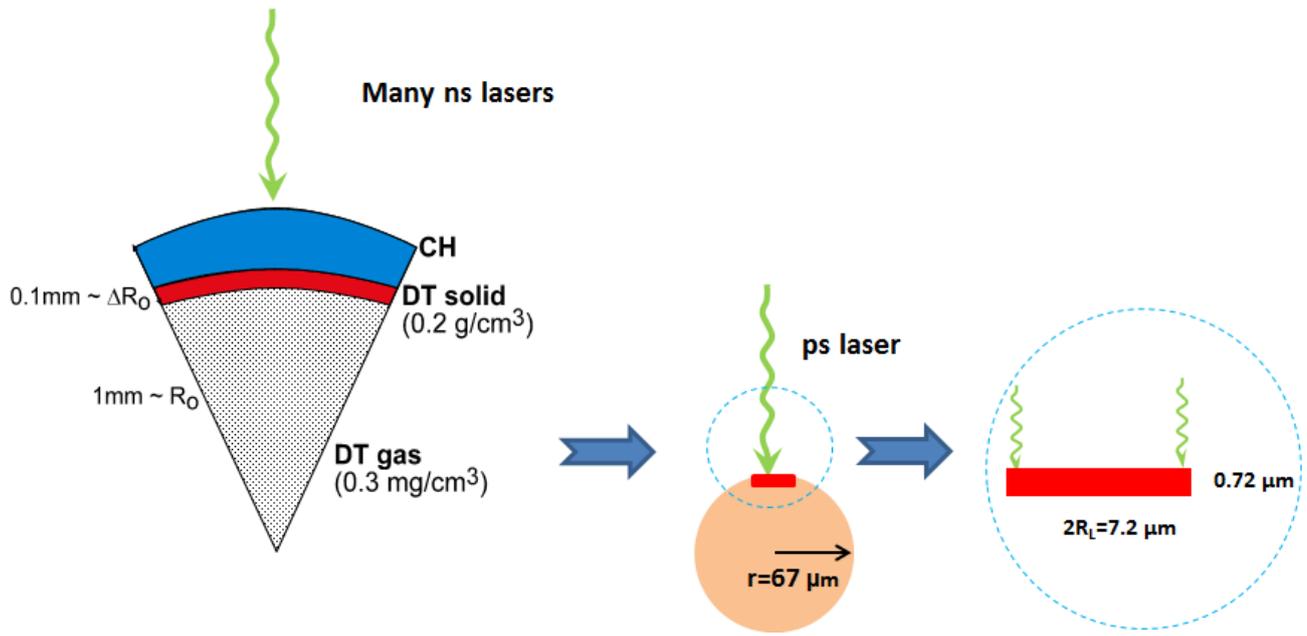